\documentclass[journal,twocolumn,10pt,twoside]{IEEECOMLetter}
\normalsize
\usepackage{amsthm,cite, balance}
\usepackage{amsmath}
\usepackage{amssymb}
\usepackage{graphicx,subfigure}
\usepackage[countmax]{subfloat}
\usepackage{color}
\usepackage{array}
\usepackage{url}
\usepackage{setspace}
\DeclareGraphicsExtensions{{.eps}{.pdf}{.jpeg}}
\begin{document}
\title{Series Representation of Modified Bessel Functions   and Its Application in AF Cooperative systems}

\author{Mehdi M. Molu\\mehdi.molu@york.ac.uk
}
\markboth{}%
{}
\maketitle

\sloppy
\begin{abstract}
Using fractional-calculus mathematics, a
novel approach is introduced to rewrite modified Bessel functions  in series
form using simple elementary functions.  Then, a statistical
characterization of the total receive-SNR at the destination,
corresponding to the source-relay-destination and the source-destination link SNR, is provided for a general relaying scenario, in which the destination exploits a 
maximum ratio combining (MRC) receiver.
 Using the novel statistical model for
the total receive SNR at the destination, accurate and simple
analytical expressions for the outage probability, the bit error
probability and the ergodic capacity are obtained.
\end{abstract}
\newtheorem{thm}{Theorem}
\newtheorem{cor}{Corollary}
\newtheorem{lem}{Lemma}
 \section{Introduction}
\label{Sec:Introduction}
For an analytical investigation of AF cooperative communication systems with a 
maximum-ratio-combining (MRC) at the destination, a statistical model
of the total receive signal-to-noise ratio, $\textit{SNR}_\text{tot}$,
is required, which is equivalent to the sum of SNRs corresponding to
the Source-Destination (S-D) and the Source-Relay-Destination (S-R-D)
link SNRs. To the best of our knowledge, a
theoretical statistical model of $\textit{SNR}_\text{tot}$ is yet
unknown (but will be presented in this paper); numerous studies,
however, consider the problem of finding a statistical model of
$\textit{SNR}_\text{srd}$. There are two main trends in the
literature: \emph{either} a single-branch system model is assumed, in
which the receiver operates only on the relay transmission and, hence,
no MRC is employed at the destination~(e.g., see  
\cite{HaAl:2003}), \emph{or} upper and/or lower
performance bounds are considered assuming an MRC receiver at the
destination~(e.g., see  \cite{FaBe:2008-j}).

To cope with the computations involving complicated mathematical
functions, a possible practical solution is to use an equivalent series
representation of the functions
(e.g.\cite{Be:1990}). In fact, the
appearance of the modified Bessel function of the second kind,
$K_\nu(\cdot)$, in the PDF of $\textit{SNR}_\text{srd}$ is the main
source of trouble in AF-related calculations. We aim to substitute
$K_\nu(\cdot)$ with an equivalent series representation in this paper.
However, as there is no appropriate and simple series representation of
$K_\nu(\cdot)$ available in the literature, motivated by \cite{Ma:2001}, we derive a novel series
representation of $K_\nu(\cdot)$ in terms of simple elementary
functions (such as $x^n \mathrm{e}^{-x}$) using fractional calculus
mathematics.  With this result, the complex statistical model of
$\textit{SNR}_\text{srd}$ turns out to be simple and easily
tractable. Thereafter, the Cumulative Distribution Function (CDF) and
the Probability Density Function (PDF) of $\textit{SNR}_\text{tot}$
will be derived. 

The remainder of the paper is organized as follows: in
Section \ref{Sec:System and Channel Model} the system model is
introduced. In Section \ref{Sec:Series K} the fractional-calculus
method is exploited to derive an equivalent series representation of
$K_\nu(x)$. Based on the results derived in Section \ref{Sec:Series
K}, the PDF and CDF of $\textit{SNR}_\text{tot}$ are derived in
Section \ref{Sec:Distribution of Equivalent Channel Power}, and in
Section \ref{Sec:Performance} novel closed-form expressions are
provided for some performance measures of variable-gain AF
cooperative systems. Conclusion remarks are provided in Section \ref{Sec:Conclusion}.

\section{System Model}
\label{Sec:System and Channel Model}

We consider a two-hop variable-gain AF cooperative system. The source (S) sends data to
the destination (D) by the help of an intermediate relay node (R). The
destination ``hears'' both the source and the relay transmissions and
employs a maximum-ratio-combining (MRC) receiver to jointly exploit
all information available at the destination. Also, motivated by practical hardware constraints, it is assumed 
that the relay operates in half-duplex mode.

The SNR at the output of the MRC receiver is ${\textit{SNR}_\text{tot}=\gamma(\vert h_\text{sd}\vert ^2 + \vert h_\text{srd}\vert ^2) = \gamma \vert h_\text{eq}\vert ^2  }$ where ${\gamma}$ is the transmit power at the source node. $\vert h_\text{sd}\vert ^2$ and  ${|h_\text{srd}|^2=\frac{|h_\text{sr}|^2|h_\text{rd}|^2}{|h_\text{sr}|^2 + |h_\text{rd}|^2 + 1/\gamma }}$ represent equivalent S-D and S-R-D channel powers, respectively.  $|h_\text{sd}|^2$, $|h_\text{sr}|^2$ and $|h_\text{rd}|^2$  are distributed exponentially with parameter $\lambda_\text{sd}$, $\lambda_\text{sr}$ and $\lambda_\text{rd}$, respectively. The CDF of $\vert h_\text{srd}\vert ^2$ is
\begin{eqnarray}
\label{eq:CDF}
F_{\vert h_\text{srd}\vert ^2}(x)=1- 2\zeta \mathrm{e}^{-\lambda_\text{S}x}
 K_1(2\zeta)
\end{eqnarray}
where $\zeta = \sqrt{\lambda_\text{P} x(x+1/\gamma)}$ and $K_\nu(\cdot)$ the modified Bessel function of the second kind
and $\nu$-th  order,  ${\lambda_\text{P}\doteq\lambda_\text{sr}\lambda_\text{rd}}$ and 
${\lambda_\text{S}\doteq\lambda_\text{sr}+\lambda_\text{rd}}$ . A proof of (\ref{eq:CDF}) can be found in\cite{MoGo2013VTC} and \cite{MoGo:2013Asilomar}.
The PDF of $\vert h_\text{srd}\vert ^2$ is
\begin{eqnarray}
\label{eq:PDF}
f_{|h_\text{srd}|^2}(x)=2 \mathrm{e}^{-\lambda_\text{S}x}\Big(\lambda_\text{P}(2x+\frac{1}{\gamma})K_0(2\zeta) +\lambda_\text{S}\zeta K_1(2\zeta)\Big).
\end{eqnarray} 
The results in (\ref{eq:PDF}) and (\ref{eq:CDF}) do not, easily, lend 
themselves to further mathematical calculations (e.g. integration) as
modified Bessel functions $K_\nu(\cdot)$ appear. Hence, no statistical
model for $\textit{SNR}_\text{tot}$ is available in the literature so
far.

In the next section, an equivalent representation of
$K_\nu(\cdot)$ is derived that is based on a series-representation
involving simple mathematical functions of the form
$x^n\mathrm{e}^{-x}$. This novel equivalent representation of
$K_\nu(\cdot)$ is then used instead of the $K_0(\cdot)$ and $K_1(\cdot)$ functions appeared in (\ref{eq:PDF}) and (\ref{eq:CDF}). It will be clear in the next sections that this approach paves the way for further theoretical analysis of AF
relaying systems.

\section{Equivalent Series Representation of Modified Bessel Functions of Second Kind}
\label{Sec:Series K}
\begin{figure}
\begin{center}
        \includegraphics[width=0.37\textwidth, height=0.35\textwidth]{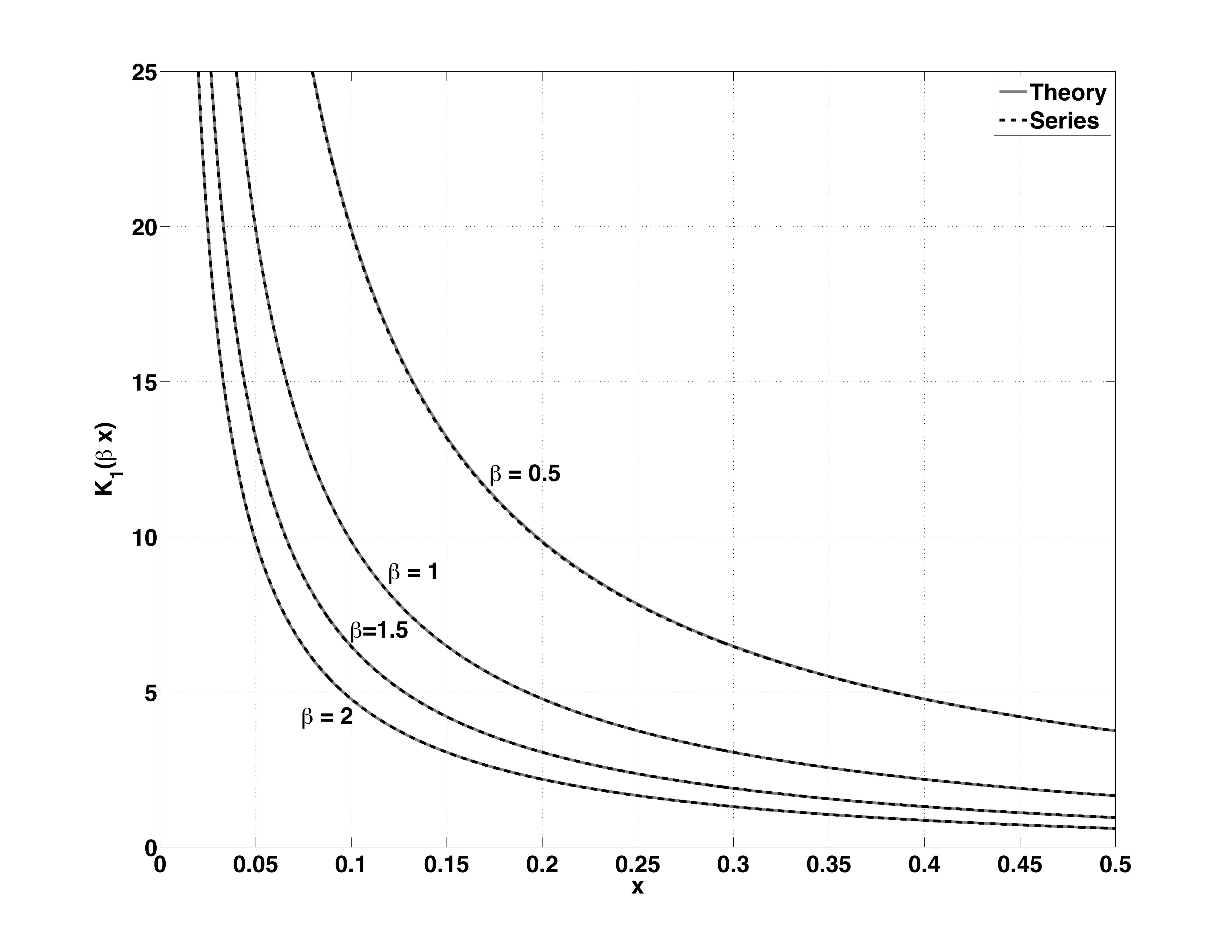}
\end{center} 
\caption{$K_1(\beta x)$ vs.~series representation of $K_1(\beta x)$ with $k=2$ in (\ref{eq:K+Trun}).}
\label{fig:bessel_vs_series}
\end{figure}

The mathematical concept of integration and differentiation of
arbitrary (non-integer) order is called ``fractional calculus'';
foundations of the theory are discussed,
e.g.,~in \cite{GoMa:1997}. It will be used below to derive a simple novel
equivalent representation of $K_\nu(\beta x)$.
\\
\textbf{Theorem.}
\textit{Equivalent Representation of $K_\nu(\beta x)$}

A modified Bessel function, $K_\nu(\beta x)$, of the second kind and
$\nu$-th order, with $\nu>0$, can be
represented by the infinite series
\begin{equation} 
\label{eq:theorem2} 
K_\nu(\beta x)
= \mathrm{e}^{-\beta x}  \: \sum\limits_{n=0}^\infty \sum\limits_{i=0}^n \Lambda(\nu,n,i) 
\cdot (\beta x)^{i-\nu}  \; ,
\end{equation} 
with the coefficients
\begin{equation} 
\label{eq:theorem2B} 
\Lambda(\nu,n,i) \doteq  
 \frac{(-1)^i\: \sqrt{\pi}\: \Gamma(2\nu)\: \Gamma(\frac{1}{2}+n-\nu) \: L(n,i)}
{2^{\nu-i} \: \Gamma(\frac{1}{2}-\nu) \: \Gamma(\frac{1}{2}+n+\nu) \: n!}
%
\end{equation} 
that involve the Lah numbers (e.g.,~\cite{Co:1974})
\begin{equation} 
\label{eq:LahNumber}
  L(n,i) \doteq  \binom{n-1}{i-1}\frac{n!}{i!} \quad \text{for} \quad n,i>0 \: ,
\end{equation} 
and the conventions $L(0,0)\doteq 1$; $L(n,0)\doteq 0$.

\begin{proof}
Let $s$ be a real non-negative number, i.e., $s>0$ and
$s \in \mathbb{C}$. Let $f(x)$ be continuous on $x\in[0,\infty)$ and
integrable on any finite subinterval of $x\geqslant 0$. Then the
Riemann-Liouville operator (e.g.,~\cite{GoMa:1997}) of fractional
integration is defined
as \begin{eqnarray} \label{eq:frac_integ_general} I^{s}\left\lbrace
f(x)\right\rbrace \doteq \frac{1}{\Gamma(s)}\int_0^x
(x-t)^{s-1}f(t)dt \; .  \end{eqnarray} On the other hand,
from \cite[3.471.4]{GrRy:2007} we
have \begin{eqnarray} \label{eq:table of integral_equation}
\int_0^x (x-t)^{s-1} t^{-2s}\mathrm{e}^{-\beta /t}dt = \frac{\Gamma(s) \beta^{\frac{1}{2}-s}}{\sqrt{\pi x}}\mathrm{e}^\frac{-\beta}{2x}K_{s-\frac{1}{2}}(\frac{\beta}{2x}) \:.
 \end{eqnarray} 
Assuming $f(t)=t^{-2s}\mathrm{e}^{-\beta /t}$, the two
integrals in (\ref{eq:frac_integ_general}) and (\ref{eq:table of
integral_equation}) are identical: this motivates the novel approach to 
derive an equivalent expression for $K_\nu(\beta x)$ by use of 
fractional integration.

It follows from
(\ref{eq:frac_integ_general}) and (\ref{eq:table of
integral_equation}) that
 \begin{eqnarray}
  \label{eq:fra==bessel}
I^{s}\left\lbrace  x^{-2s}\mathrm{e}^{-\beta /x}  \right\rbrace  =  \frac{\beta^{\frac{1}{2}-s}}{\sqrt{\pi x}}\mathrm{e}^\frac{-\beta}{2x}K_{s-\frac{1}{2}}(\frac{\beta}{2x}) \; .
 \end{eqnarray}
The Leibniz rule for the Riemann-Liouville operator is
\begin{eqnarray}
\label{eq:leibniz}
I^{s}\left\lbrace h(x)g(x)\right\rbrace =\sum\limits_{n=0}^{\infty}\frac{(-1)^n \Gamma(n+s)}{n! \Gamma(s)}I^{s+n}\left\lbrace h\right\rbrace D^{n}\left\lbrace g\right\rbrace
\end{eqnarray}
where $n$ is a non-negative integer, $s+n$ is a non-negative
fractional number and $D^n\doteq\frac{d^n}{dx^n}$. By solving
$I^{s+n}\left\lbrace h(x)\right\rbrace$ for $h(x)=x^{-2s}$ and
$D^{n}\left\lbrace g(x)\right\rbrace$ for $g(x)=\mathrm{e}^{-\beta/x}$, the
equivalent Bessel model (\ref{eq:theorem2}) will be derived.  

Let $h(x)=x^p$, then according to (\ref{eq:frac_integ_general})
\begin{eqnarray}
\label{eq:frac_power_general}
I^{\alpha}\{x^p\} &=& \frac{1}{\Gamma(\alpha)}\int_0^x (x-t)^{\alpha -1} t^p dt, \hspace{5.7mm}(\alpha >0) \\
\label{eq:frac_power_general_4}
&=&\frac{\Gamma(1+p)}{\Gamma(1+p+\alpha)}x^{p+\alpha} \: .
\end{eqnarray}
 with $\Gamma(\cdot)$ the
 Gamma-function.
Suppose that $p=-2s$ and $\alpha=s+n$, then 
\begin{equation}
\label{eq:frac_power}
I^{(s+n)}\left\lbrace x^{-2s}\right\rbrace =  \frac{\Gamma(1-2s)}{\Gamma(1-s+n)}x^{n-s} \: .
\end{equation}

With $g(x) = \mathrm{e}^{-\beta/x}$ in (\ref{eq:leibniz}),  $D^n \left\lbrace \mathrm{e}^{-\beta/x} \right\rbrace $ can be found to equal
\begin{eqnarray}
\label{eq:Der(e^(-Bx))_final}
D^n \left\lbrace \mathrm{e}^{-\beta/x} \right\rbrace =
 \mathrm{e}^{-\beta/x}
\frac{(-1)^{n}}{x^n}\sum_{i=0}^{n} (-1)^{i} L(n,i)(\beta/x)^i \: ,
\end{eqnarray}
with $L(n,i)$ defined in (\ref{eq:LahNumber}); this
result is taken from \cite{DaMaSpTa:2012}.

By substituting (\ref{eq:frac_power}) and
(\ref{eq:Der(e^(-Bx))_final}) into (\ref{eq:fra==bessel}) and
(\ref{eq:leibniz}) it is straightforward to
obtain 
\begin{eqnarray}
\label{eq:serie=bessel}
&&\hspace*{-7mm}\frac{\beta^{\frac{1}{2}-s}}{\sqrt{\pi x}}\mathrm{e}^{-\frac{\beta}{2x}}K_{s-\frac{1}{2}}(\frac{\beta}{2x})=
\sum\limits_{n=0}^\infty \frac{(-1)^n \Gamma(n+s)\Gamma(1-2s)}{n!\: \Gamma(s)\Gamma(1+n-s)} x^{n-s}
\nonumber\\
&& \hspace*{17mm}\times\mathrm{e}^{-\beta/x}
\frac{(-1)^{n}}{x^n}\sum_{i=0}^{n} (-1)^{i} L(n,i)(\beta/x)^i .
\end{eqnarray}
Changing the variable
$x\to \frac{1}{2x}$, assuming $1-2s=2\nu$ and exploiting
$K_{-\nu}=K_{\nu}$, the result is the infinite series in \eqref{eq:theorem2}.
\end{proof}

It should be made clear that the above representation of $K_\nu(\beta
x)$ is \textit{not} valid for $\nu=\left\lbrace
0, \frac{1}{2},\frac{3}{2},\cdots\right\rbrace$. That is because
$\Gamma(2\nu)$ and $\Gamma(\frac{1}{2}+n-\nu)$ in (\ref{eq:theorem2B})
diverge to $\pm\infty$. However, for the case of $\nu=0$, one can
compute $K_0(\beta x)$ using the equivalent representation of
$K_1(\beta x)$ and $K_2(\beta x)$ by $K_\nu(\beta x)= K_{\nu-2}( \beta
x)+ \frac{2(\nu -1)}{\beta x} K_{\nu-1}(\beta x)$ that is obtained
from \cite[10.38.4]{OlLoBoCl:2010}. 

\subsection*{Finite Series Representation of $K_\nu(\beta x)$} 
The equivalent representation of $K_\nu(\beta x)$ may significantly
simplify computations involving $K_\nu(\beta x)$, as the series in
(\ref{eq:theorem2}) contains the variable $x$ only in the simple
function-template $x^{i-\nu} \mathrm{e}^{-\beta x}$ that can, e.g., be easily
integrated. The series representation contains, however, an infinite
number of terms that can't be computed in practical applications.

Fortunately, the series representation of $K_\nu(\beta x)$ is rather
accurate for a finite number of terms as defined as follows:
\begin{eqnarray}
\label{eq:K+Trun}
K_\nu(\beta x)
= \mathrm{e}^{-\beta x} \sum\limits_{n=0}^{k} \sum\limits_{i=0}^n \Lambda(\nu,n,i) \cdot
(\beta x)^{i-\nu} +  \epsilon
\end{eqnarray}
with
\begin{eqnarray}
\label{eq:Tr.Err}
\epsilon &=& \mathrm{e}^{-\beta x} \sum\limits_{n=k+1}^\infty \sum\limits_{i=0}^n \Lambda(\nu,n,i) \cdot (\beta x)^{i-\nu}.
\end{eqnarray}
The first term on the right-hand side of (\ref{eq:K+Trun}) represents
the actual function to approximate $K_\nu(\beta x)$, and $\epsilon$
represents the truncation error that can be neglected. Fig.~\ref{fig:bessel_vs_series}
illustrates numerical values of the finite series representation of $K_1(\beta x)$ (with
$k=2$ in (\ref{eq:K+Trun}))  for various values of $\beta$
(dashed lines) and also the theoretical fully accurate values of
$K_1(\beta x)$ (solid lines). It is clear from the figure that the finite
series for $K_\nu(\beta x)$ with only $k=2$ merges with theoretical $K_\nu(\beta x)$ with high accuracy.

\paragraph*{Convenient Representation for Practical Use of the Truncated Series} 
For practical use, it is convenient to re-write (\ref{eq:K+Trun}) such
that one of the sum-operators is included in the series
coefficients. For this, the inner sum over $i$ in (\ref{eq:K+Trun}) is
evaluated row-wise, with the sum-index $n$ counting the rows. This
structure is then summed up column-wise and the result can be written
as
\begin{eqnarray}
\label{eq:K+Trun2}
K_\nu(\beta x)
& = & \frac{\mathrm{e}^{-\beta x}}{(\beta x)^\nu} \: \sum\limits_{q=0}^{k} 
\underbrace{\Big(\sum\limits_{l=q}^k \Lambda(\nu,l,q) \Big)}_{\displaystyle\doteq a_{\nu,k,q}} 
\cdot (\beta x)^{q} + \epsilon \: .
\end{eqnarray}
As long as $k$ is limited, the above series representation
(\ref{eq:K+Trun2}) can always be used to replace (\ref{eq:K+Trun})
without any convergence problems but still the truncation error
$\epsilon$ can be made arbitrarily small. Numerical values for the
coefficients $a_{\nu,k,q}$ are given in
Table \ref{tab:PowerSeriesCoeffs} for the first-order ($\nu =1$)
modified Bessel function of the second kind $K_1(\cdot)$; it should be
noted that the coefficient with $q$-index $1$ is always found to equal
$a_{1,k,1} \doteq \frac{2k}{2k+1}$.

\begin{center}
\begin{table*}[ht]
\normalsize
\hfill{}
\begin{tabular}{|c||c|c|c|c|c|c|c|c|c|c|c|}
\hline
&$a_{1,k,0}$ &  $a_{1,k,1}$    &  $a_{1,k,2}$  &$a_{1,k,3}$  & $a_{1,k,4}$ & $a_{1,k,5}$ & $a_{1,k,6}$ & $a_{1,k,7}$ & $a_{1,k,8}$ & $a_{1,k,9}$ & $a_{1,k,10}$  \\\hline\hline
$k = 2$  &$1$ & $4/5$ & $-0.1333$   \\\cline{1-7}  
$k = 5$   & $1$ & $10/11$ &$-0.4237$ & $0.1824$     &$-0.0375$ & $ 2.693$ \\
&&&&&&$\times10^{-3}$\\\cline{1-12}  
$k = 10$ &$1$ &  $20/21$ & $-0.7047$ & $0.7239$ &$-0.5000$ &$0.2111$&$-5.415$&$8.375$&$-7.55$ &$3.619$ & $-7.0724$\\
&&&&&&&$\times10^{-2}$ &$\times10^{-3}$ &$\times10^{-4}$&$\times10^{-5}$&$\times10^{-7}$
\\\hline
\end{tabular}
\hfill{}
\caption{Coefficients $a_{\nu,k,q}$ in (\ref{eq:K+Trun2}) for $\nu =1$ with four digits of accuracy }
\label{tab:PowerSeriesCoeffs}
\hrulefill
\vspace*{4pt}
\end{table*}
\end{center}
\section{Distribution of Equivalent Channel Power}
\label{Sec:Distribution of Equivalent Channel Power}
Assuming an MRC receiver at the destination,
the total receive SNR is the sum of SNRs corresponding to the S-D and the S-R-D links, i.e., $\textit{SNR}_\text{tot} = \textit{SNR}_\text{sd} + \textit{SNR}_\text{srd} = 
 \gamma (\vert h_\text{sd}\vert^2 +\vert h_\text{srd}\vert^2)$.
The equivalent channel power-gain can be written as
$
\vert h_\text{eq}\vert^2 \doteq\vert
h_\text{srd}\vert^2+\vert h_\text{sd}\vert^2
$,
and so, for the CDF of $\vert h_\text{eq}\vert^2$ we obtain
\begin{eqnarray}
\label{eq:ProbOut_Bessel}
F_{\vert h_\text{eq}\vert^2}(x)&=&\mathbb{P}\left(\vert h_\text{sd}\vert^2 +\vert h_\text{srd}\vert^2 \leq x \right)\nonumber  \\
&=&\lambda_\text{sd}\mathrm{e}^{-\lambda_\text{sd}x}\int_0^x \mathrm{e}^{\lambda_\text{sd}u}\cdot F_{\vert h_\text{srd}\vert^2}(u)du \: ,
\end{eqnarray}
with $F_{\vert h_\text{srd}\vert^2}(\cdot)$ derived in (\ref{eq:CDF}). For simplicity we will restrict
calculations below to the high ``transmit-SNR'' regime but it will be
demonstrated, by simulations~(e.g.,
Fig.~\ref{fig:Erg_C})
that this is justified because it leads to very accurate numerical
results, even in the low-SNR region.

By plugging (\ref{eq:CDF})
into (\ref{eq:ProbOut_Bessel}), and assuming ``high SNR'', (\ref{eq:ProbOut_Bessel}) simplifies to
\begin{eqnarray}
\label{eq:ProbOut_Bessel_high_SNR}
F_{\vert h_\text{eq}\vert^2}(r)&=&1-\mathrm{e}^{-\lambda_\text{sd}r}-2\lambda_\text{sd}\sqrt{\lambda_\text{sr}\lambda_\text{rd}}\mathrm{e}^{-\lambda_\text{sd}r} \\
&&\times \int_0^r x\mathrm{e}^{-(\lambda_\text{sr}+\lambda_\text{rd}-\lambda_\text{sd})x}K_1(  2\sqrt{\lambda_\text{sr}\lambda_\text{rd}} x) dx \: . \nonumber
\end{eqnarray}
\begin{figure}
\begin{center}
        \includegraphics[width=0.35\textwidth]{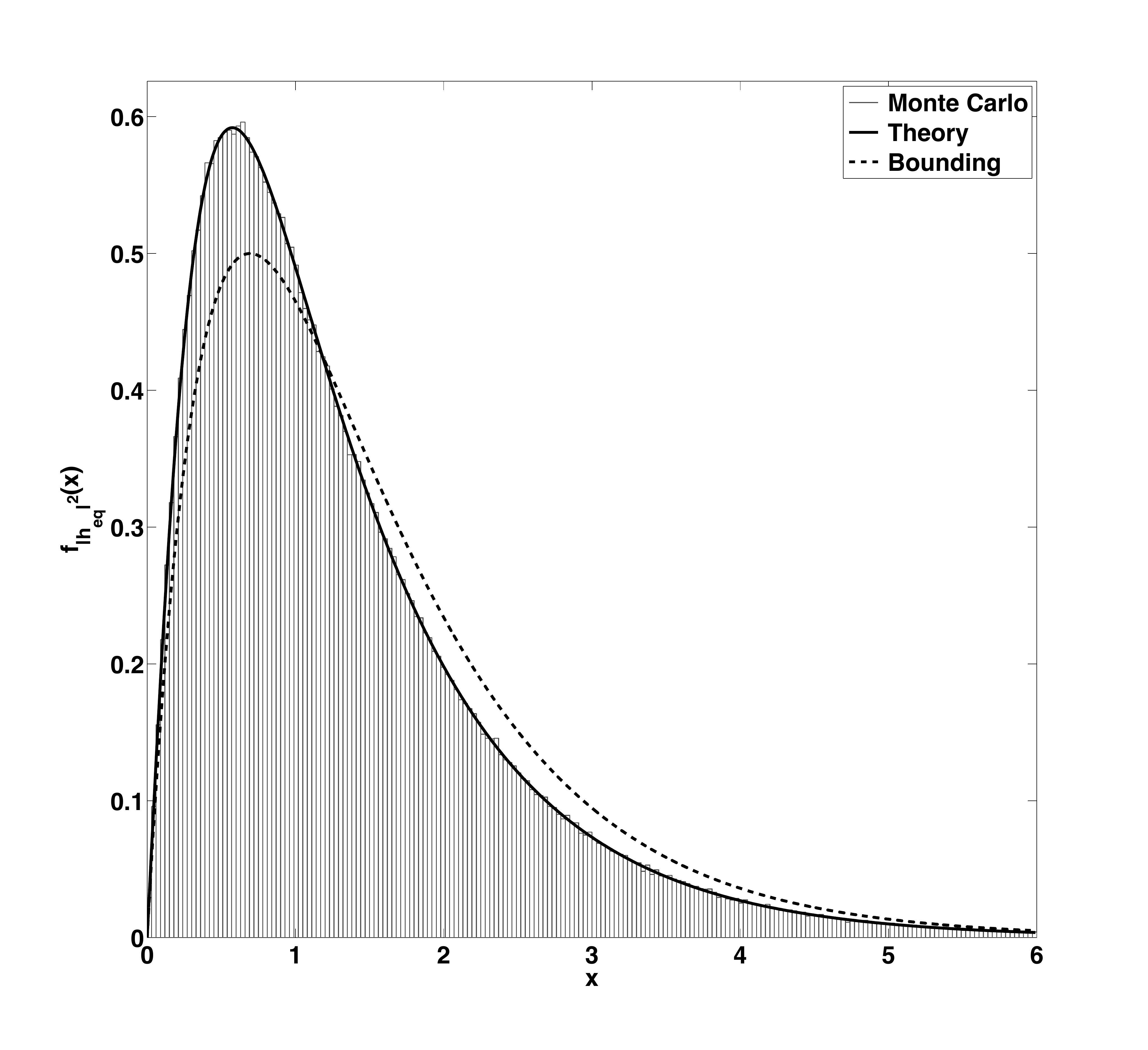}
\end{center} 
\caption{PDF $f_{\vert h\text{eq} \vert^2}(x)$ using the theoretical model 
derived in (\ref{eq:PDF_Heq_final}); Monte Carlo simulations and the
bounding technique proposed in the literature
(e.g.~\cite{IkAh:2007}), all with
$\lambda_\text{sr}=\lambda_\text{rd}=1$.}
\label{fig:PDF_Heq}
\end{figure} 
The integral in (\ref{eq:ProbOut_Bessel_high_SNR}) is non-trivial and
does not seem to have closed-form solution. However, using  
the results from Section \ref{Sec:Series K},
the integral can be rewritten as follows 
\begin{eqnarray}
\label{eq:eta}
\eta&\doteq&\int_0^x 2\sqrt{\lambda_\text{sr}\lambda_\text{rd}}u\mathrm{e}^{-(\lambda_\text{sr}+\lambda_\text{rd}-\lambda_\text{sd})u}K_1(\; 2\sqrt{\lambda_\text{sr}\lambda_\text{rd}} u) dx\nonumber\\
&\approx&\sum\limits_{q=0}^{k} \frac{(2\sqrt{\lambda_\text{sr}\lambda_\text{rd}})^q q!\: a_{1,k,q} }{(\lambda_{\text{srd}} -\lambda_{\text{sd}})^{q+1}}\nonumber\\
&&\hspace*{10mm}\times \big(1-\sum\limits_{c=0}^{q} \frac{(\lambda_{\text{srd}} -\lambda_{\text{sd}})^{c}}{c!}x^c\mathrm{e}^{-(\lambda_{\text{srd}}-\lambda_{\text{sd}}) x}\big)
\end{eqnarray}
where $\lambda_{\text{srd}} =
(\sqrt{\lambda_\text{sr}}+ \sqrt{\lambda_\text{rd}})^2$. The second step is obtained by using the series representation
of $K_1(2\sqrt{\lambda_\text{sr}\lambda_\text{rd}} x)$ derived in
(\ref{eq:K+Trun2}). As the series is truncated (for
$k$ limited), this is an approximation, indicated by the use of
``$\approx$'' instead of strict equality; the truncation error can,
however, be made arbitrarily small by choosing proper $k$.
By substituting (\ref{eq:eta}) into (\ref{eq:ProbOut_Bessel_high_SNR})
it is straightforward to obtain $F_{\vert h_\text{eq}\vert^2}(x)$ as
\begin{equation}
\label{eq:Prob Out final}
F_{\vert{h_\text{eq}}\vert^2}(x)\approx 1-\mathcal{A}\mathrm{e}^{-\lambda_\text{sd} x}+\sum\limits_{q=0}^{k}\sum\limits_{c=0}^{q}\mathcal{B}x^c\mathrm{e}^{-\lambda_{\text{srd}}x}
\end{equation}
where coefficients $\mathcal{A}$ and $\mathcal{B}$ are independent of $x$, defined as
\begin{eqnarray}
\label{eq:A an B}
\mathcal{A} &\doteq& 1+ \sum\limits_{q=0}^{k}\frac{\lambda_\text{sd}(2\sqrt{\lambda_\text{sr}\lambda_\text{rd}})^q q!\: a_{1,k,q}}{(\lambda_{\text{srd}} -\lambda_{\text{sd}})^{q+1}}\\
\mathcal{B} &\doteq& \frac{\lambda_\text{sd}(2\sqrt{\lambda_\text{sr}\lambda_\text{rd}})^q q!\: a_{1,k,q}}{c! \: (\lambda_{\text{srd}} -\lambda_{\text{sd}})^{q-c+1}} \: .
\end{eqnarray}
The PDF of $\vert{h_\text{eq}}\vert^2$ is the derivative of
$F_{\vert{h_\text{eq}}\vert^2}(x)$ in (\ref{eq:Prob Out final}) w.r.t $x$, which is easy to calculate as the series representation involves simple elementary functions only:
\begin{equation}
\label{eq:PDF_Heq_final}
f_{\vert{h_\text{eq}}\vert^2}(x)\approx \mathcal{A}\lambda_\text{sd}\mathrm{e}^{-\lambda_\text{sd} x}+\sum\limits_{q=0}^{k}\sum\limits_{c=0}^{q} \mathcal{B}(cx^{c-1}-\lambda_{\text{srd}}x^c)\mathrm{e}^{-\lambda_{\text{srd}}x} \: .
\end{equation}
Fig.~\ref{fig:PDF_Heq} illustrates the accuracy of $f_{\vert{h_\text{eq}}\vert^2}(x)$
using the expression derived in (\ref{eq:PDF_Heq_final}) (solid line)
by a comparison with histogram-results obtained from Monte Carlo
simulations.

In the literature 
(e.g.~\cite{IkAh:2007}), when
considering relaying systems with MRC receiver at the destination, a
method is proposed for estimating the statistics of
$\textit{SNR}_\text{srd}$ based on the bounding technique
$\textit{SNR}_\text{srd} \doteq \min
(\textit{SNR}_\text{sr},\textit{SNR}_\text{rd})$. Consequently,
$\textit{SNR}_\text{srd}$ has as an exponential distribution with
parameter $\lambda_\text{sr}+\lambda_\text{rd}$. Note that, although
this method greatly simplifies the calculations by avoiding modified
Bessel functions in the formulations, accuracy is sacrificed:
Fig.~\ref{fig:PDF_Heq} (dashed line) shows the result for
$f_{\vert{h_\text{eq}}\vert^2}(x)$ when using the bounding method as well. It
is clear that the approach presented in this paper results in more accurate solutions in comparison with  traditional bounding techniques. 

\section{Performance Analysis}
\label{Sec:Performance}
The CDF in  \eqref{eq:CDF} corresponds to the outage probability. Due to lack of space no illustration  is provided for outage probability but with the accuracy of the PDF in Fig.~\ref{fig:PDF_Heq}, the accuracy of outage probability is ensured as well.

Given the PDF of $\vert{h_\text{eq}}\vert^2$ in \ref{eq:PDF_Heq_final}, bit error probability (BEP) can be obtained as ${p_\text{b}
= \frac{1}{2}\int_0^\infty \operatorname{erfc}(\sqrt{\gamma x}) f_{\vert{h_\text{eq}}\vert^2}(x)dx}$.
From \cite[7.1.19]{AbSt:2000}, $\frac{d}{dx} \operatorname{erfc}(\sqrt{\gamma x}) = -\sqrt{\frac{\gamma}{\pi}} x^{-1/2} \mathrm{e}^{-\gamma x}$. Then, using integration by parts, $p_\text{b}$ can be written according to
\begin{eqnarray}
\label{eq:BEP final}
p_\text{b}=\frac{1}{2}\Big(1-\sqrt{\frac{\gamma \mathcal{A}^2}{\gamma+\lambda_\text{sd}}}+\sqrt{\frac{\gamma}{\pi}}\sum\limits_{q=0}^{k}\sum\limits_{c=0}^{q}\frac{\mathcal{B}\Gamma(c+\frac{1}{2})}{(\gamma + \lambda_\text{srd})^{c+\frac{1}{2}}}\Big).
\end{eqnarray}
\begin{figure}[t]
\begin{center}
        \includegraphics[width=0.35\textwidth]{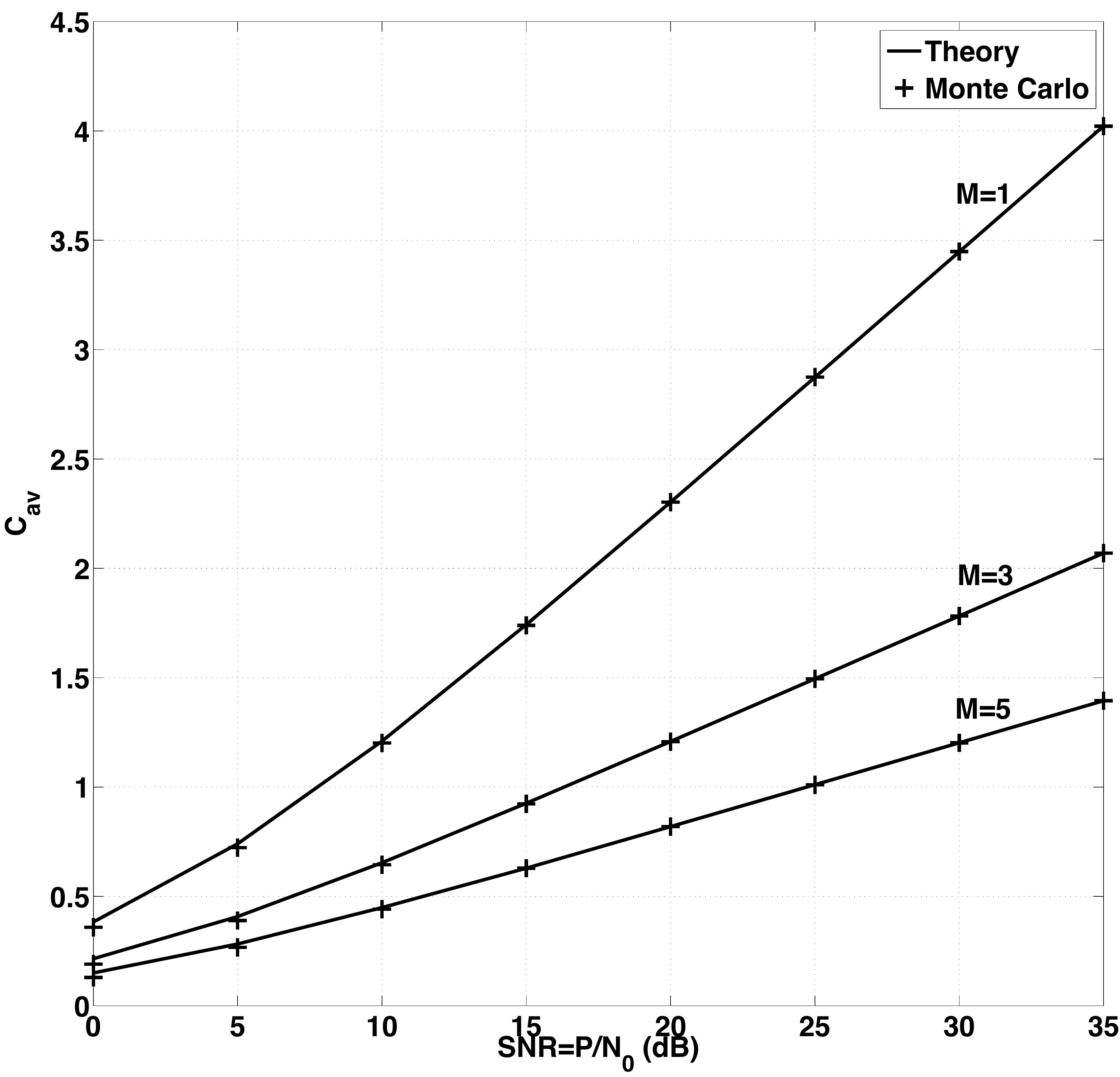}
\end{center} 
\caption{Ergodic Capacity (nats/sec/Hz) of the system with multiple relays and equal transmit power from the relays.}
\label{fig:Erg_C}
\end{figure} 

Ergodic capacity of an AF cooperative system can be obtained using  the definition of capacity as ${C_{av} = \frac{1}{2} \int_0^{\infty}\log(1+\gamma x)f_{\vert h_\text{eq}\vert^2}(x) \mathrm{d}x}$, that can easily be solved in closed-form by substituting $f_{\vert h_\text{eq}\vert^2}(x)$ from \eqref{eq:PDF_Heq_final} and exploiting   \cite[4.222.8]{GrRy:2007}. Due to space limit, the closed form expression is not provided in this letter; however, Fig.\ref{fig:Erg_C} illustrates the Ergodic capacity obtained from the proposed approach. Also, the proposed approach has been examined for multi relay scenarios as well. The excellent agreement between Monte-Carlo Simulations and closed-form solution proves the correctness of the results. 
\section{Conclusions}
\label{Sec:Conclusion}
 Based on fractional calculus mathematics, a novel power series
representation of $K_\nu(\cdot)$ is introduced that has simple
elementary functions of the form $x^n\mathrm{e}^{-x}$ as its basic
components. Based on that the performance of AF cooperative systems is investigated and accurate results are obtained.

\balance
\bibliography{S_Ref.bib}

\begin{thebibliography}{10}

\bibitem{HaAl:2003}
M.~O. Hasna and M.~S. Alouini, ``End-to-end performance of transmission systems
  with relays over {R}ayleigh fading channels,'' {\em IEEE Transactions on
  Wireless Communications}, vol.~2, November 2003.

\bibitem{FaBe:2008-j}
G.~Farhadi and N.~Beaulieu, ``On the ergodic capacity of wireless relaying
  systems over {R}ayleigh fading channels,'' {\em IEEE Transactions on Wireless
  Communications}, vol.~7, pp.~4462--4467, Nov. 2008.

\bibitem{Be:1990}
N.~Beaulieu, ``An infinite series for the computation of the complementary
  probability distribution function of a sum of independent random variables
  and its application to the sum of {R}ayleigh random variables,'' {\em IEEE
  Transactions on Communications}, vol.~38, pp.~1463--1474, Sept. 1990.

\bibitem{Ma:2001}
K.~Maslanka, ``Series representation of the modified {B}essel functions.''
  \url{http://arxiv.org/abs/math-ph/0104018v1}, April 2001.

\bibitem{MoGo2013VTC}
M.~M.~Molu and N.~Goertz, ``An analytical approach to the outage probability of
  amplify-and-forward relaying with an {MRC} receiver,'' in {\em Proceedings
  Int. Conf. on Vehicular Technology}, June 2013.

\bibitem{MoGo:2013Asilomar}
M.~M.~Molu and N.~Goertz, ``Performance analysis of amplify-and-forward
  relaying using fractional calculus,'' in {\em Proceedings of Asilomar
  Conference on Signals, Systems, and Computers}, June 2012.

\bibitem{GoMa:1997}
R.~Gorenflo and F.~Mainardi, {\em Fractional calculus: Integral and
  differential equations of fractional order}, vol.~378 of CISM Courses and
  Lectures.
\newblock Springer-Verlag, 1997.

\bibitem{Co:1974}
L.~Comtet, {\em Advanced Combinatorics: The Art of Finite and Infinite
  Expansions}.
\newblock D. Reidel Publishing Company, 1974.

\bibitem{GrRy:2007}
I.~Gradshteyn and I.~Ryzhik, {\em Table of Integrals, Series, and Products}.
\newblock Academic Press, 7th~ed., 2007.

\bibitem{DaMaSpTa:2012}
S.~Daboul, J.~Mangaldan, M.~Spivey, and P.~Taylor, ``The {L}ah numbers and the
  nth derivative of $e^{1/x}$.''
  \url{http://math.pugetsound.edu/~mspivey/Exp.pdf} (accessed 6 August 2012).

\bibitem{OlLoBoCl:2010}
F.~W.~J. Oliver, D.~W. Lozier, R.~F. Boisvert, and C.~W. Clark, {\em NIST
  Handbook of Mathematical Functions}.
\newblock Cambridge Press, July 2010.

\bibitem{IkAh:2007}
S.~Ikki and M.~Ahmed, ``Performance analysis of cooperative diversity wireless
  networks over {N}akagami-m fading channel,'' {\em IEEE Communications
  Letters}, vol.~11, pp.~334--336, Apr. 2007.

\bibitem{AbSt:2000}
Abramowitz and Stegun, {\em Handbook of Mathematical Functions}.
\newblock Dover Publications, June 1965.

\end{thebibliography}

\end{document}